\documentclass[12pt,thmsa]{article}
\usepackage{amsfonts}
\usepackage{amsmath}
\usepackage{sw20lart}

\input tcilatex
\begin{document}

\author{Nick Laskin\thanks{\textit{E-mail address}: nlaskin@rocketmail.com}}
\title{\textbf{Stretched Coherent States}}
\date{TopQuark Inc.\\
Toronto, ON, M6P 2P2\\
Canada}
\maketitle

\begin{abstract}
The fundamental properties of recently introduced stretched coherent states
are investigated. \ It has been shown that stretched coherent states retain
the fundamental properties of standard coherent states and generalize the
resolution of unity, or completeness condition, and the probability
distribution that $n$ photons are in a stretched coherent state.

The stretched displacement and stretched squeezing operators are introduced
and the multiplication law for stretched displacement operator is
established. The results of the action of the stretched displacement and
stretched squeezing operators on the vacuum and the Fock states are
presented.

Stretched squeezed stretched coherent states and stretched squeezed
stretched displaced number states are introduced and their properties are
studied.

The inner product of two quantum mechanical vectors was defined in terms of
their stretched coherent state representations, and functional Hilbert space
was introduced.


\end{abstract}

\section{Introduction}

\textit{Standard coherent states} were discovered by Schr\"{o}dinger \cite%
{Schrodinger} in his search for quantum states in whose representation the
diagonal matrix element of the evolution operator of a quantum mechanical
oscillator exhibits the same temporal behavior as a classical mechanical
oscillator. The term "\textit{coherent states}"\ was coined by Glauber \cite%
{Glauber}, who introduced these states as superpositions of Fock states of
the quantized electromagnetic field that are not modified by the action of
photon annihilation operators. In other words, Glauber found that the
quantum state of a coherent quantized field has to be an eigenvector of the
boson-annihilation operator with complex eigenvalue.

In this note we explore the fundamental properties of recently introduced
stretched coherent states which generalize Glauber's coherent states
framework. We have applied the concept of stretched coherent states to
design stretched squeezed stretched coherent states, and stretched squeezed
stretched displaced number states.

The motivation for introducing and developing these states is twofold.
First, we expand the foundations of quantum optics: quantum coherent and
squeezed coherent states, as well as coherent displaced number states.
Second, presented generalizations of the well-known fundamental concepts
open up new horizons for researchers working in the field of quantum optics
to search for manifestations and applications of the new fundamentals in
optical experiments. The first step in this direction has been made in the
seminal paper by Longhi \cite{Longhi}, where the optical realization of
fractional quantum mechanics \cite{LaskinFQM} was proposed based on
transverse light dynamics in aspherical optical cavities. As a laser
implementation of the fractional quantum harmonic oscillator it has been
found that dual Airy beams can be selectively generated under off-axis
longitudinal pumping. Another interesting realization of fractional quantum
mechanics was implemented in \cite{Malomed}, based on similarity between the
standard Schr\"{o}dinger equation and paraxial wave-propagation equation in
optics. The authors of the paper \cite{Malomed} proposed a protocol that
uses the transverse dynamics of light in aspherical optical cavities
designed in the $4f$ configuration. In the paraxial approximation and
without taking into account losses on optical elements and diffraction, the
output transverse modes correspond to the eigenfunctions of the fractional
Schr\"{o}dinger equation \cite{LaskinfSch}.

It was shown in \cite{Longhi} and \cite{Malomed} that the use of fractional
models in the field of optics allows to manage diffraction of light and
design novel signal-processing schemes and beam solutions.

Therefore, we hope that this note will initiate a search for whether
stretched quantum coherent states play the same role in fractional quantum
mechanics as standard coherent states in quantum mechanics, and will
stimulate new developments and experiments in the field of quantum optics.

We present the proof that stretched coherent states are indeed generalized
coherent states and explore their properties. The stretch displacement
operator and newly introduced stretched squeezing operator have been
studied. It has been shown that stretched coherent states retain the
fundamental properties of standard coherent states and generalize the
resolution of unity or completeness condition, the probability distribution
that $n$ photons are in a stretched coherent state, the multiplication law
for stretched displacement operator, and the results of action of the
stretched displacement and stretched squeezing operators on the vacuum and
the Fock states.

The inner product of two quantum mechanical vectors was defined in terms of
their stretched coherent state representations, and functional Hilbert space
was introduced.

The presented new concepts of quantum optics include the parameters $\sigma $%
, $0<\sigma \leq 1$ and $\upsilon $, $0<\upsilon \leq 1$. In the limiting
case, when $\sigma =1$ and $\upsilon =1$, all our new results turn into
known equations of the theory of standard coherent states.

\section{Stretched coherent states}

The \textit{stretched coherent states} $|\varsigma >_{\sigma }$, which are a
generalization of standard coherent states, were introduced as follows \cite%
{Laskin},

\begin{equation}
|\varsigma >_{\sigma }=\exp (-\frac{|\varsigma |^{2\sigma }}{2}%
)\sum\limits_{n=0}^{\infty }\frac{\varsigma ^{\sigma n}}{\sqrt{n!}}%
|n>,\qquad 0<\sigma \leq 1,  \label{eq7.17}
\end{equation}

and the adjoint states $_{\sigma }<\varsigma |$%
\begin{equation}
_{\sigma }<\varsigma |=\exp (-\frac{|\varsigma |^{2\sigma }}{2}%
)\sum\limits_{n=0}^{\infty }\frac{(\varsigma ^{\sigma \ast })^{n}}{\sqrt{n!}}%
<n|,\qquad 0<\sigma \leq 1,  \label{eq7.18}
\end{equation}

where a complex number $\varsigma $ stands for labelling the stretched
coherent states, vector $|n>$ is an eigenvector of the photon number
operator $\overset{\wedge }{n}=a^{+}a$, \ 
\begin{equation}
\overset{\wedge }{n}|n>=n|n>,  \label{eq.7.18_1}
\end{equation}%
$<n|m>=\delta _{n,m}$, the operators $a^{+}$ and $a$ are photon field
creation and annihilation operators that satisfy the Bose-Einstein
commutation relation $[a,a^{+}]=aa^{+}-a^{+}a=\QTR{sl}{1}$, and $%
[a^{+},a^{+}]=0$, $[a,a]=0$. The action of the operators $a^{+}$ and $\ a$
on the number state $|n>$ reads

{}%
\begin{equation}
a^{+}|n>=\sqrt{n+1}|n+1>\qquad \text{and}\qquad a|n>=\sqrt{n}|n-1>.
\label{eq7.18a}
\end{equation}

It is easy to see that stretched quantum coherent state $|\varsigma
>_{\sigma }$ is eigenstate of photon field annihilation operator $a$. \
Indeed, we have

\begin{equation}
a|\varsigma >_{\sigma }=\exp (-\frac{|\varsigma |^{2\sigma }}{2}%
)\sum\limits_{n=0}^{\infty }\frac{\varsigma ^{\sigma n}}{\sqrt{n!}}\sqrt{n}%
|n-1>=  \label{eq7.20}
\end{equation}

\begin{equation*}
\exp (-\frac{|\varsigma |^{2\sigma }}{2})\sum\limits_{n=0}^{\infty }\frac{%
\varsigma ^{\sigma (n+1)}}{\sqrt{n!}}|n>=\varsigma ^{\sigma }|\varsigma
>_{\sigma },
\end{equation*}

that is

\begin{equation}
a|\varsigma >_{\sigma }=\varsigma ^{\sigma }|\varsigma >_{\sigma },
\label{eq.7.20a}
\end{equation}

which shows that the eigenvalue of the photon field annihilation operator $a$
is $\varsigma ^{\sigma }$. Therefore, we have

\begin{equation*}
_{\sigma }<\varsigma |a|\varsigma >_{\sigma }=\varsigma ^{\sigma },\qquad 
\text{and\qquad }_{\sigma }<\varsigma |a^{+}|\varsigma >_{\sigma }=\varsigma
^{\sigma \ast }.
\end{equation*}

Using the following representation for $|n>$

\begin{equation}
|n>=\frac{(a^{+})^{n}}{\sqrt{n!}}|0>,  \label{eq7.21}
\end{equation}

where $|0>$ is the vacuum state with $\varsigma =0$, we obtain the
alternative expression for the stretched quantum coherent states

\begin{equation}
|\varsigma >_{\sigma }=\exp (-\frac{|\varsigma |^{2\sigma }}{2}%
)\sum\limits_{n=0}^{\infty }\frac{(\varsigma ^{\sigma }a^{+})^{n}}{\sqrt{n!}}%
|0>,\qquad 0<\sigma \leq 1.  \label{eq7.22}
\end{equation}

\section{Fundamental properties of the stretched coherent states}

Let us answer the question whether the states $|\varsigma >_{\sigma }$ are
generalized coherent states. Quantum mechanical states are generalized
coherent states if they \cite{Klauder2}:

(i) are parameterized continuously and normalized;

(ii) admit a resolution of unity with a positive weight function;

(iii) provide temporal stability, that is, a coherent state that evolves
over time belongs to the family of coherent states.

To prove (i), we note that the stretched coherent states $|\varsigma
>_{\sigma }$ introduced by Eq.(\ref{eq7.17}) are evidently parametrized
continuously by their label $\varsigma $ which is a complex number $%
\varsigma =\xi +i\eta $, with $\xi =\mathrm{Re}\varsigma $ and $\eta =%
\mathrm{Im}\varsigma $. The square of the modulus $|<n|\varsigma >_{\sigma
}|^{2}$ of projection of the stretched coherent state $|\varsigma >_{\sigma
} $ onto the number state $|n>$ gives us the probability $P_{\sigma
}(n,\varsigma )$ that $n$ photons will be found in a coherent state $%
|\varsigma >_{\sigma },$

\begin{equation}
P_{\sigma }(n,\varsigma )=|<n|\varsigma >_{\sigma }|^{2}=\frac{|\varsigma
|^{2\sigma n}}{n!}\exp (-|\varsigma |^{2\sigma }),\qquad 0<\sigma \leq 1.
\label{eq8.1}
\end{equation}

Therefore, the stretched quantum coherent states $|\varsigma >_{\sigma }$
are normalized due to the normalization condition for the probability
distribution $P_{\sigma }(n,\varsigma )$,

\begin{equation}
\sum\limits_{n=0}^{\infty }|<n|\varsigma >_{\sigma
}|^{2}=\sum\limits_{n=0}^{\infty }P_{\sigma }(n,\varsigma )=1.  \label{eq8.2}
\end{equation}

To prove (ii), that is, the coherent states $|\varsigma >_{\sigma }$ admit a
resolution of unity with a positive weight function, we introduce a function 
$W_{\sigma }(|\varsigma |^{2})>0$ which obeys the equation

\begin{equation}
\int\limits_{%
\mathbb{C}
}d^{2}\varsigma |\varsigma >_{\sigma }W_{\sigma }(|\varsigma |^{2})_{\sigma
}<\varsigma |=I,  \label{eq8.3}
\end{equation}

where $d^{2}\varsigma =d(\mathrm{Re}\varsigma )d(\mathrm{Im}\varsigma )$ and
the integration extends over the entire complex plane $%
\mathbb{C}
$. This equation with yet unknown function $W_{\sigma }(|\varsigma |^{2})$
is the resolution of unity for stretched coherent states $|\varsigma
>_{\sigma }$. Introducing new integration variables $r$ and $\varphi $ by $%
\varsigma =re^{i\varphi }$, $d^{2}\varsigma =rdrd\varphi $ and making use of
Eqs.(\ref{eq7.17}) and (\ref{eq7.18}) yield

\begin{equation*}
\int\limits_{%
\mathbb{C}
}d^{2}\varsigma |\varsigma >_{\sigma }W_{\sigma }(|\varsigma |^{2})_{\sigma
}<\varsigma |=
\end{equation*}

\begin{equation}
\underset{\Phi \rightarrow \infty }{\lim }\frac{1}{2\Phi }%
\sum\limits_{n,m=0}^{\infty }\int\limits_{-\Phi }^{\Phi }d\varphi e^{i\sigma
(n-m)\varphi }\int\limits_{0}^{\infty }drr^{(n+m)\sigma +1}\frac{W_{\sigma
}(r^{2})}{\sqrt{n!m!}}\exp (-r^{2\sigma })|n><m|=I,  \label{eq8.4}
\end{equation}

where we used Klauder's "covering space formulation" ansatz \cite{Klauder2}
to perform the integration over $d\varphi $. Due to

\begin{equation}
\underset{\Phi \rightarrow \infty }{\lim }\frac{1}{2\Phi }\int\limits_{-\Phi
}^{\Phi }d\varphi e^{i\sigma (n-m)\varphi }=\delta _{m,n},  \label{eq8.5}
\end{equation}

we come to the following equation to find the function $W_{\sigma }(r^{2})$

\begin{equation}
\sum\limits_{n=0}^{\infty }\frac{1}{n!}\int\limits_{0}^{\infty
}drr^{2n\sigma +1}W_{\sigma }(r^{2})\exp (-r^{2\sigma })|n><n|=I.
\label{eq8.6}
\end{equation}

Hence we conclude that if a positive function $W_{\sigma }(r^{2})$ satisfies
the equation

\begin{equation}
\frac{1}{n!}\int\limits_{0}^{\infty }drr^{2n\sigma +1}W_{\sigma }(r^{2})\exp
(-r^{2\sigma })=1,  \label{eq8.7a}
\end{equation}

then due to completeness of orthonormal vectors $|n>$

\begin{equation}
\sum\limits_{n=0}^{\infty }|n><n|=I,  \label{eq8.8}
\end{equation}

the resolution of unity expressed by Eq.(\ref{eq8.3}) will hold. It is easy
to see, that $W_{\sigma }(r^{2})$ must be $W_{\sigma }(r^{2})=2\sigma
r^{2(\sigma -1)}$ to satisfy Eq.(\ref{eq8.7a})$,$ or

\begin{equation}
W_{\sigma }(|\varsigma |^{2})=2\sigma |\varsigma |^{2(\sigma -1)}.
\label{eq8.9a}
\end{equation}

Therefore, the resolution of unity is

\begin{equation}
2\sigma \int\limits_{%
\mathbb{C}
}d^{2}\varsigma |\varsigma >_{\sigma }|\varsigma |^{2(\sigma -1)}{}_{\sigma
}<\varsigma |=I,  \label{eq8.11}
\end{equation}

which can be seen as a completeness condition for the stretched coherent
states $|\varsigma >_{\sigma }$.

To prove (iii), we note that if $|n>$ is an eigenvector of the Hamiltonian
operator $\overset{\wedge }{H}=\hbar \omega \overset{\wedge }{n}=\hbar
\omega a^{+}a$, where $\hbar $ is Planck's constant, then the time evolution
operator $\exp (-iHt/\hbar )$ results

\begin{equation*}
\exp (-i\overset{\wedge }{H}t/\hbar )|n>=e^{-i\omega nt}|n>.
\end{equation*}

In other words, the time evolution of $|n>$ results in appearance of the
phase factor only. Let's consider time evolution of the stretched coherent
state $|\varsigma >_{\sigma }$ defined by Eq.(\ref{eq7.17}). Since the
stretched coherent state is not an eigenstate of $\overset{\wedge }{H}$, one
would expect it to evolve to other states over time. However, we see that

\begin{equation}
\exp (-i\overset{\wedge }{H}t/\hbar )|\varsigma >_{\sigma }=\exp (-\frac{%
|\varsigma |^{2\sigma }}{2})\sum\limits_{n=0}^{\infty }\frac{\varsigma
^{\sigma n}}{\sqrt{n!}}e^{-i\omega nt}|n>=|e^{-\frac{i\omega t}{\sigma }%
}\varsigma >_{\sigma },  \label{eq8.12}
\end{equation}

which is just another coherent state belonging to a complex number $%
\varsigma e^{-\frac{i\omega t}{\sigma }}$. Hence, the time evolution of the
stretched coherent state $|\varsigma >_{\sigma }$ remains within the family
of the coherent states $|\varsigma >_{\sigma }$. The property embodied in
Eq.(\ref{eq8.12}) is called the temporal stability of coherent states $%
|\varsigma >_{\sigma }$ under the action of the time evolution operator.

Thus, we conclude that stretched quantum coherent states $|\varsigma
>_{\sigma }$ satisfy Klauder's criteria (i) - (iii)\ for generalized
coherent states \cite{Klauder2}.

\subsection{Mandel parameter}

The probability $P_{\sigma }(n,\varsigma )$ that the field represented by
stretched coherent state \TEXTsymbol{\vert}$\varsigma >_{\sigma }$ is
occupied by $n$ photons is given by Eq.(\ref{eq8.1}). The mean number of
photons in the quantum state $|\varsigma >_{\sigma }$ is

\begin{equation}
_{\sigma }<\varsigma |\overset{\wedge }{n}|\varsigma >_{\sigma }=_{\sigma
}<\varsigma |a^{+}a|\varsigma >_{\sigma }=\sum\limits_{n=0}^{\infty
}nP_{\sigma }(n,\varsigma )=|\varsigma |^{2\sigma },  \label{eq8.13}
\end{equation}

and the second order moment of the number of photons in the quantum state $%
|\varsigma >_{\sigma }$ is

\begin{equation}
_{\sigma }<\varsigma |\overset{\wedge }{n}^{2}|\varsigma >_{\sigma
}=_{\sigma }<\varsigma |(a^{+}a)^{2}|\varsigma >_{\sigma
}=\sum\limits_{n=0}^{\infty }n^{2}P_{\sigma }(n,\varsigma )=|\varsigma
|^{2\sigma }+|\varsigma |^{4\sigma },  \label{eq8.14}
\end{equation}

where $a^{+}$ and $a$ are photon field creation and annihilation operators.
These equations allow us to calculate the Mandel parameter \cite{Mandel}
using stretched coherent states. For one-mode quantum fields represented by
the stretched coherent states the Mandel parameter $Q_{\sigma }$ is given by

\begin{equation}
Q_{\sigma }=\frac{_{\sigma }<\varsigma |(a^{+}a)^{2}|\varsigma >_{\sigma
}-(_{\sigma }<\varsigma |a^{+}a|\varsigma >_{\sigma })^{2}}{_{\sigma
}<\varsigma |a^{+}a|\varsigma >_{\sigma }}-1.  \label{eq8.15}
\end{equation}

Taking into account Eqs.(\ref{eq8.13}) and (\ref{eq8.14}) we conclude that $%
Q_{\sigma }=0$. In other words, stretched quantum coherent states obey
Poisson statistics.

\section{Stretched displacement operator}

We introduce the \textit{stretched displacement operator} $D_{\sigma
}(\varsigma )$ as follows

\begin{equation}
D_{\sigma }(\varsigma )=\exp \{\varsigma ^{\sigma }a^{+}-\varsigma ^{\sigma
\ast }a\},\qquad 0<\sigma \leq 1,  \label{eq7.24}
\end{equation}

where $\varsigma $ is a complex number, $a^{+}$ and $a$ are photon field
creation and annihilation operators.

When $\sigma =1$ the operator $D_{\sigma }(\varsigma )|_{\sigma =1}$ becomes
the well-known displacement operator $D(\varsigma )$ for the standard
coherent states \cite{Schrodinger}, \cite{Glauber}, \cite{Klauder},

\begin{equation}
D(\varsigma )=D_{\sigma }(\varsigma )|_{\sigma =1}=\exp \{\varsigma
a^{+}-\varsigma ^{\ast }a\}.  \label{eq7.24_1}
\end{equation}

The stretched displacement operator $D_{\sigma }(\varsigma )$ is an unitary
operator, i.e.

\begin{equation}
D_{\sigma }^{+}(\varsigma )D_{\sigma }(\varsigma )=1,  \label{eq7.24a}
\end{equation}

where the sign "$+"$ stands for Hermitian conjugation of the operator.

Using the Baker-Campbell-Hausdorff formula for operators $\widehat{A}$ and $%
\widehat{B}$,

\begin{equation}
e^{\widehat{A}+\widehat{B}}=e^{\widehat{A}}e^{\widehat{B}}e^{-\frac{1}{2}[%
\widehat{A},\widehat{B}]},  \label{eq7.24b}
\end{equation}

such that%
\begin{equation}
\lbrack \widehat{A},[\widehat{A},\widehat{B}]]=[\widehat{B},[\widehat{A},%
\widehat{B}]]=0,  \label{eq.7.24c}
\end{equation}

we come to alternative representations for the stretched displacement
operator $D_{\sigma }(\varsigma )$,

\begin{equation}
D_{\sigma }(\varsigma )=\exp (-\frac{|\varsigma |^{2\sigma }}{2})\exp
(\varsigma ^{\sigma }a^{+})\exp (-\varsigma ^{\sigma \ast }a),
\label{eq7.25a}
\end{equation}

or

\begin{equation}
D_{\sigma }(\varsigma )=\exp (\frac{|\varsigma |^{2\sigma }}{2})\exp
(-\varsigma ^{\sigma \ast }a)\exp (\varsigma ^{\sigma }a^{+}).
\label{eq7.25b}
\end{equation}

To show that

\begin{equation}
|\varsigma >_{\sigma }=D_{\sigma }(\varsigma )|0>,  \label{eq7.26}
\end{equation}

we perform the following chain of transformations

\begin{equation*}
D_{\sigma }(\varsigma )|0>=\exp (-\frac{|\varsigma |^{2\sigma }}{2})\exp
(\varsigma ^{\sigma }a^{+})\exp (-\varsigma ^{\sigma \ast }a)|0>=
\end{equation*}

\begin{equation*}
\exp (-\frac{|\varsigma |^{2\sigma }}{2})\exp (\varsigma ^{\sigma
}a^{+})|0>=\exp (-\frac{|\varsigma |^{2\sigma }}{2})\sum\limits_{n=0}^{%
\infty }\frac{\varsigma ^{\sigma n}(a^{+})^{n}}{n!}|0>=
\end{equation*}

\begin{equation*}
\exp (-\frac{|\varsigma |^{2\sigma }}{2})\sum\limits_{n=0}^{\infty }\frac{%
\varsigma ^{\sigma n}}{\sqrt{n!}}|n>=|\varsigma >_{\sigma },
\end{equation*}

where Eq.(\ref{eq7.21}) was used. Therefore, operator $D_{\sigma }(\varsigma
)$ generates stretched coherent state $|\varsigma >_{\sigma }$ from the
vacuum state $|0>.$

It is easy to see that the following equations hold

\begin{equation}
\lbrack a,D_{\sigma }(\varsigma )]=\varsigma ^{\sigma }D_{\sigma }(\varsigma
),  \label{eq7.27}
\end{equation}

and

\begin{equation}
D_{\sigma }^{+}(\varsigma )aD_{\sigma }(\varsigma )=a+\varsigma ^{\sigma
},\qquad \qquad D_{\sigma }(\varsigma )aD_{\sigma }^{+}(\varsigma
)=a-\varsigma ^{\sigma }.  \label{eq7.28}
\end{equation}

\textit{Multiplication law} for the stretched displacement operators can be
established using the Baker-Campbell-Hausdorff formula Eq.(\ref{eq7.24b}),

\begin{equation}
D_{\sigma }(\varsigma )D_{\sigma }(\eta )=D(\varsigma ^{\sigma }+\eta
^{\sigma })\exp \{\frac{1}{2}(\varsigma ^{\sigma }\eta ^{\sigma \ast
}-\varsigma ^{\sigma \ast }\eta ^{\sigma })\},  \label{eq7.31}
\end{equation}

with $D(\varsigma ^{\sigma }+\eta ^{\sigma })$ being the well-known
displacement operator defined by Eq.(\ref{eq7.24_1}).

With help of Eqs.(\ref{eq7.31}) and (\ref{eq7.24b}) it can be shown that the
matrix element $<m|D_{\sigma }(\varsigma )|n>$ of the stretched displacement
operator in the number state representation $|n>$ is expressed as

\begin{equation}
<m|D_{\sigma }(\varsigma )|n>=\sqrt{\frac{n!}{m!}}\varsigma ^{\sigma
(m-n)}\exp (-\frac{|\varsigma |^{2\sigma }}{2})L_{n}^{(m-n)}(|\varsigma
|^{2\sigma }),  \label{eq7.32}
\end{equation}

here $L_{n}^{(m-n)}(x)$ are the associated Laguerre polynomials, the
generating function of which has form (see, Eq.(19), page 189, in \cite%
{Bateman2})%
\begin{equation}
\sum\limits_{n=0}^{\infty }L_{n}^{(m-n)}(x)y^{n}=e^{-xy}(1+y)^{m},\qquad
|y|<1.  \label{eq7.33}
\end{equation}

The diagonal matrix element $<n|D_{\sigma }(\varsigma )|n>$ is

\begin{equation}
<n|D_{\sigma }(\varsigma )|n>=\exp (-\frac{|\varsigma |^{2\sigma }}{2}%
)L_{n}(|\varsigma |^{2\sigma }),  \label{eq7.34}
\end{equation}

where $L_{n}(x)$ is the Laguerre polynomial of order $n$, related to the
associated Laguerre polynomial $L_{n}^{(m-n)}(x)$ as follows, $%
L_{n}(x)=L_{n}^{(0)}(x)$.

Finally note, that due to Eqs.(\ref{eq7.24a}), (\ref{eq7.26}) and the
orthonormality of the vectors $|n>$, the scalar product $_{\sigma }<\eta
|\varsigma >_{\sigma }$ can be expressed as

\begin{equation}
_{\sigma }<\eta |\varsigma >_{\sigma }=\exp \{-\frac{|\eta |^{2\sigma }}{2}-%
\frac{|\varsigma |^{2\sigma }}{2}+\eta ^{\sigma \ast }\varsigma ^{\sigma }\},
\label{eq7.23}
\end{equation}

which is the \textit{overcompleteness} relation for the stretched quantum
coherent states $|\varsigma >_{\sigma }$.

\section{Stretched squeezed stretched coherent states}

Let us introduce \textit{stretched squeezed stretched coherent states }$%
|\varsigma ,\xi >_{\sigma ,\upsilon }$ as follows,

\begin{equation}
|\varsigma ,\xi >_{\sigma ,\upsilon }=D_{\sigma }(\varsigma )S_{\upsilon
}(\xi )|0>,\qquad 0<\sigma \leq 1,\quad 0<\upsilon \leq 1,  \label{eq9.1a}
\end{equation}

with the stretched displacement operator $D_{\sigma }(\varsigma )$ defined
by Eq.(\ref{eq7.24}) and the \textit{stretched squeezing operator} $%
S_{\upsilon }(\xi )$, which we introduce as follows,

\begin{equation}
S_{\upsilon }(\xi )=\exp \{\frac{1}{2}\xi ^{\upsilon \ast }a^{2}-\frac{1}{2}%
\xi ^{\upsilon }a^{+2}\},\qquad 0<\upsilon \leq 1,  \label{eq9.1}
\end{equation}

where the squeeze parameter $\xi =\rho \exp (i\theta )$ is an arbitrary
complex number, and $a^{+}$ and $a$ are photon field creation and
annihilation operators.

Quantum states $|\varsigma ,\xi >_{\sigma ,\upsilon }$ represent the new
family of coherent states, which includes the following members.

The well-known squeezed coherent states,

\begin{equation}
|\varsigma ,\xi >=D_{\sigma }(\varsigma )|_{\sigma =1}S_{\upsilon }(\xi
)|_{\upsilon =1}|0>=D(\varsigma )S(\xi )|0>,\qquad \sigma =1,\quad \upsilon
=1,  \label{eq9.1b}
\end{equation}

where $D(\varsigma )$ is defined by Eq.(\ref{eq7.24_1})\ and $S(\xi )$ is
given by

\begin{equation}
S(\xi )=\exp \{\frac{1}{2}\xi ^{\ast }a^{2}-\frac{1}{2}\xi a^{+2}\}.
\label{eq9.1c}
\end{equation}

The new \textit{stretched squeezed coherent states},

\begin{equation}
|\varsigma ,\xi >_{\upsilon }=D(\varsigma )|_{\sigma =1}S_{\upsilon }(\xi
)|0>=D(\varsigma )S_{\upsilon }(\xi )|0>,\qquad \sigma =1,\quad 0<\upsilon
\leq 1.  \label{eq9.1d}
\end{equation}

The new \textit{squeezed stretched coherent states},

\begin{equation}
|\varsigma ,\xi >_{\sigma }=D_{\sigma }(\varsigma )S_{\upsilon }(\xi
)|_{\upsilon =1}|0>=D_{\sigma }(\varsigma )S(\xi )|0>,\qquad 0<\sigma \leq
1,\quad \upsilon =1.  \label{eq9.1e}
\end{equation}

The stretched squeezing operator $S_{\upsilon }(\xi )$ is unitary operator

\begin{equation}
S_{\upsilon }^{+}(\xi )S_{\upsilon }(\xi )=1.  \label{eq9.11}
\end{equation}

It is easy to see that the following transformations hold for the creation $%
a^{+}$ and annihilation $a$ operators

\begin{equation}
S_{\upsilon }^{+}(\xi )a^{+}S_{\upsilon }(\xi )=a^{+}\cosh \rho ^{\upsilon
}-ae^{-i\upsilon \theta }\sinh \rho ^{\upsilon },  \label{eq9.12}
\end{equation}

and

\begin{equation}
S_{\upsilon }^{+}(\xi )aS_{\upsilon }(\xi )=a\cosh \rho ^{\upsilon
}-a^{+}e^{i\upsilon \theta }\sinh \rho ^{\upsilon },  \label{eq9.13}
\end{equation}

Let's show, as an example, the technique to calculate the expectation $%
_{\sigma }<\varsigma ,\xi |a|\varsigma ,\xi >_{\sigma }$ of annihilation
operator in the stretched squeezed stretched coherent states\textit{\ }$%
|\varsigma ,\xi >_{\sigma ,\upsilon }$ basis. Using the definition Eq.(\ref%
{eq9.1a}) we write

\begin{equation}
_{\sigma ,\upsilon }<\varsigma ,\xi |a|\varsigma ,\xi >_{\sigma ,\upsilon
}=<0|S_{\upsilon }^{+}(\xi )D_{\sigma }^{+}(\varsigma )aD_{\sigma
}(\varsigma )S_{\upsilon }(\xi )|0>=  \label{eq9.4_1}
\end{equation}

\begin{equation*}
<0|S_{\upsilon }^{+}(\xi )(a+\varsigma ^{\sigma })S_{\upsilon }(\xi )|0>,
\end{equation*}

where the last transition took into account the first of Eq.(\ref{eq7.28}).
Further, using Eq.(\ref{eq9.13}) we get

\begin{equation}
_{\sigma ,\upsilon }<\varsigma ,\xi |a|\varsigma ,\xi >_{\sigma ,\upsilon
}=\varsigma ^{\sigma }.  \label{eq9.4_2}
\end{equation}

Similarly, one can calculate the following expectations

\begin{equation}
_{\sigma ,\upsilon }<\varsigma ,\xi |a^{2}|\varsigma ,\xi >_{\sigma
,\upsilon }=|\varsigma |^{2\sigma }-e^{2i\upsilon \theta }\sinh \rho
^{\upsilon }\cosh \rho ^{\upsilon },  \label{eq9.5}
\end{equation}

and

\begin{equation}
_{\sigma ,\upsilon }<\varsigma ,\xi |a^{+}a|\varsigma ,\xi >_{\sigma
,\upsilon }=|\varsigma |^{2\sigma }+\sinh ^{2}\rho ^{\upsilon }.
\label{eq9.6}
\end{equation}

\subsection{Stretched squeezed stretched displaced number states}

Using operators $S(\xi )$\ and $D_{\sigma }(\varsigma )$ defined by Eqs.(\ref%
{eq9.1}) and (\ref{eq7.24})\ respectively, we introduce the \textit{%
stretched squeezed stretched displaced number states},

\begin{equation}
|\varsigma ,\xi ,n>_{\sigma ,\upsilon }=D_{\sigma }(\varsigma )S_{\upsilon
}(\xi )|n>,  \label{eq.10.1}
\end{equation}

here $|n>$ in the number state or Fock state and operators $D_{\sigma
}(\varsigma )$ and $S_{\upsilon }(\xi )$ are defined by Eq.(\ref{eq7.24})
and Eq.(\ref{eq9.1}) respectively.

For $n=0$, the stretched squeezed stretched displaced number state $%
|\varsigma ,\xi ,n>_{\sigma ,\upsilon }|_{n=0}$ \ becomes the stretched
squeezed stretched coherent state $|\varsigma ,\xi >_{\sigma ,\upsilon }$
defined by Eq.(\ref{eq9.1a}).

For $\varsigma =0$, the stretched squeezed stretched displaced number state $%
|\varsigma ,\xi ,n>_{\sigma }|_{\varsigma =0}$ becomes stretched squeezed
displaced number state $|\xi ,n>_{\upsilon }$ defined by

\begin{equation}
|\xi ,n>_{\upsilon }=S_{\upsilon }(\xi )|n>.  \label{eq10.2}
\end{equation}

For $\xi =0$, the stretched squeezed stretched displaced number state $%
|\varsigma ,\xi ,n>_{\sigma }|_{\xi =0}$ becomes stretched displaced number
state $|\varsigma ,n>_{\sigma }$ defined by

\begin{equation}
|\varsigma ,n>_{\sigma }=D_{\sigma }(\varsigma )|n>.  \label{eq10.3a}
\end{equation}%
With help of Eq.(\ref{eq7.32}) the stretched displaced number state can be
expressed as

\begin{equation}
|\varsigma ,n>_{\sigma }=D_{\sigma }(\varsigma )|n>=\exp (-\frac{|\varsigma
|^{2\sigma }}{2})\sum\limits_{m=0}^{\infty }\sqrt{\frac{n!}{m!}}\varsigma
^{\sigma (m-n)}L_{n}^{(m-n)}(|\varsigma |^{2\sigma })|m>.  \label{eq10.3}
\end{equation}

The stretched displaced number state $|\varsigma ,n>_{\sigma }$ can be
expressed in terms of the stretched coherent state $|\varsigma >_{\sigma }$
introduced by Eq.(\ref{eq7.17}). Indeed, using Eq.(\ref{eq7.21}) we have

\begin{equation}
|\varsigma ,n>_{\sigma }=D_{\sigma }(\varsigma )|n>=D_{\sigma }(\varsigma )%
\frac{(a^{+})^{n}}{\sqrt{n!}}|0>=  \label{eq10.4}
\end{equation}

\begin{equation*}
D_{\sigma }(\varsigma )\frac{(a^{+})^{n}}{\sqrt{n!}}D_{\sigma
}^{+}(\varsigma )D_{\sigma }(\varsigma )|0>=D_{\sigma }(\varsigma )\frac{%
(a^{+})^{n}}{\sqrt{n!}}D_{\sigma }^{+}(\varsigma )|\varsigma >_{\sigma }.
\end{equation*}

Further, using the second of Eq.(\ref{eq7.28})\ we get

\begin{equation}
D_{\sigma }(\varsigma )\frac{(a^{+})^{n}}{\sqrt{n!}}D_{\sigma
}^{+}(\varsigma )=\frac{1}{\sqrt{n!}}\left( D_{\sigma }(\varsigma
)a^{+}D_{\sigma }^{+}(\varsigma )\right) ^{n}=\frac{1}{\sqrt{n!}}%
(a^{+}-\varsigma ^{\sigma \ast })^{n}.  \label{eq10.5}
\end{equation}

Hence, combining Eqs.(\ref{eq10.4}) and (\ref{eq10.5})\ yields

\begin{equation}
|\varsigma ,n>_{\sigma }=\frac{1}{\sqrt{n!}}(a^{+}-\varsigma ^{\sigma \ast
})^{n}|\varsigma >_{\sigma }.  \label{eq10.6}
\end{equation}

This equation initiates the introduction of the \textit{modified stretched
displacement operator }$D_{\sigma }(\alpha ,\varsigma ),$

\begin{equation}
D_{\sigma }(\alpha ,\varsigma )=\exp \{\alpha ^{\sigma }(a^{+}-\varsigma
^{\sigma \ast })-\alpha ^{\sigma \ast }(a-\varsigma ^{\sigma })\},
\label{eq10.7}
\end{equation}

and \textit{modified stretched coherent state }$|\alpha ,\varsigma >_{\sigma
}$,

\begin{equation}
|\alpha ,\varsigma >_{\sigma }=D_{\sigma }(\alpha ,\varsigma )|\varsigma
>_{\sigma }=\exp \{\alpha ^{\sigma }(a^{+}-\varsigma ^{\sigma \ast })-\alpha
^{\sigma \ast }(a-\varsigma ^{\sigma }\}|\varsigma >_{\sigma },
\label{eq10.8}
\end{equation}

or

\begin{equation}
|\alpha ,\varsigma >_{\sigma }=\exp \{\alpha ^{\sigma \ast }\varsigma
^{\sigma }-\alpha ^{\sigma }\varsigma ^{\sigma \ast }\}D_{\sigma }(\alpha
)|\varsigma >_{\sigma },  \label{eq10.9}
\end{equation}

where $D_{\sigma }(\alpha )$ is the stretched displacement operator defined
by Eq.(\ref{eq7.24}).

Let us show that the modified stretched coherent state can be expressed in
terms of the number states. From Eq.(\ref{eq10.9}) we have

\begin{equation}
|\alpha ,\varsigma >_{\sigma }=\exp \{\alpha ^{\sigma \ast }\varsigma
^{\sigma }-\alpha ^{\sigma }\varsigma ^{\sigma \ast }\}\exp (-\frac{%
|\varsigma |^{2\sigma }}{2})\sum\limits_{n=0}^{\infty }\frac{\varsigma
^{\sigma n}}{\sqrt{n!}}D_{\sigma }(\alpha )|n>=  \label{eq10.10}
\end{equation}

\begin{equation*}
\exp \{\alpha ^{\sigma \ast }\varsigma ^{\sigma }-\alpha ^{\sigma }\varsigma
^{\sigma \ast }\}\exp (-\frac{|\varsigma |^{2\sigma }}{2})\sum%
\limits_{n=0}^{\infty }\frac{\varsigma ^{\sigma n}}{\sqrt{n!}}|\alpha
,n>_{\sigma },
\end{equation*}

where we used the definition given by Eq.(\ref{eq10.2}). Then Eq.(\ref%
{eq10.3}) gives us an expression for the modified stretched coherent state $%
|\alpha ,\varsigma >_{\sigma }$ in terms of the number states $|m>,$

\begin{equation*}
|\alpha ,\varsigma >_{\sigma }=e^{\alpha ^{\sigma \ast }\varsigma ^{\sigma
}-\alpha ^{\sigma }\varsigma ^{\sigma \ast }}e^{-\frac{|\varsigma |^{2\sigma
}+|\alpha |^{2\sigma }}{2}}\sum\limits_{n,m=0}^{\infty }\frac{\varsigma
^{\sigma n}}{\sqrt{m}}\alpha ^{\sigma (m-n)}L_{n}^{(m-n)}(|\alpha |^{2\sigma
})|m>,
\end{equation*}

where $L_{n}^{(m-n)}(x)$ are the associated Laguerre polynomials \cite%
{Bateman2}.

\section{Quantum mechanical vector and operator representations based on
stretched coherent states $|\protect\varsigma >_{\protect\sigma }$}

The resolution of unity condition Eq.(\ref{eq8.3}), with $W_{\sigma
}(|\varsigma |^{2})$ given by Eq.(\ref{eq8.9a}), allows us to introduce the
inner product of two quantum mechanical vectors.

1. \textit{Inner Product} of quantum mechanical vectors\textit{\ }$|\varphi
> $\textit{\ }and\textit{\ }$|\psi >$ defined as

\begin{equation}
<\varphi |\psi >_{\sigma }=\frac{1}{\pi }\int\limits_{%
\mathbb{C}
}d^{2}\varsigma <\varphi |\varsigma >_{\sigma }W_{\sigma }(|\varsigma
|^{2})_{\sigma }<\varsigma |\psi >,  \label{eq30}
\end{equation}

where $d^{2}\varsigma =d(\mathrm{Re}\varsigma )d(\mathrm{Im}\varsigma )$ and
the integration extends over the entire complex plane $%
\mathbb{C}
$, the vector representatives are wave functions $<\varphi |\varsigma
>_{\sigma }$ and $_{\sigma }<\varsigma |\psi >$ given by 
\begin{equation}
<\varphi |\varsigma >_{\sigma }=\exp (-\frac{|\varsigma |^{2\sigma }}{2}%
)\sum\limits_{n=0}^{\infty }\frac{\varsigma ^{\sigma n}}{\sqrt{n!}}<\varphi
|n>,  \label{eq31}
\end{equation}

\begin{equation}
_{\sigma }<\varsigma |\psi >=\exp (-\frac{|\varsigma |^{2\sigma }}{2}%
)\sum\limits_{n=0}^{\infty }<n|\psi >\frac{(\varsigma ^{\sigma \ast })^{n}}{%
\sqrt{n!}}.  \label{eq32}
\end{equation}

Having the inner product, we introduce the following transformation laws.

2. \textit{Vectors Transformation Law}

\begin{equation}
_{\sigma }<\varsigma |\mathcal{A}|\psi >=\int\limits_{%
\mathbb{C}
}d^{2}\varsigma _{\sigma }^{^{\prime }}<\varsigma |\mathcal{A}|\varsigma
^{^{\prime }}>_{\sigma }W_{\sigma }(|\varsigma ^{^{\prime }}|^{2})_{\sigma
}<\varsigma ^{^{\prime }}|\psi >,  \label{eq33}
\end{equation}

where $_{\sigma }<\varsigma |\mathcal{A}|\varsigma ^{^{\prime }}>_{\sigma }$
is the matrix element of quantum mechanical operator $\mathcal{A}$.

3. \textit{Operator Transformation Law}

\begin{equation}
_{\sigma }<\varsigma |\mathcal{A}_{1}\mathcal{A}_{2}|\varsigma ^{^{\prime
}}>_{\sigma }=\int\limits_{%
\mathbb{C}
}d^{2}\varsigma _{\sigma }^{^{^{\prime \prime }}}<\varsigma |\mathcal{A}%
_{1}|\varsigma ^{^{^{\prime \prime }}}>_{\sigma }W_{\sigma }(|\varsigma
^{^{\prime \prime }}|^{2})_{\sigma }<\varsigma ^{^{\prime \prime }}|\mathcal{%
A}_{2}|\varsigma ^{^{\prime }}>_{\sigma },  \label{eq34}
\end{equation}

where $\mathcal{A}_{1}$ and $\mathcal{A}_{2}$ are two quantum mechanical
operators.

Further, the inverse map from the functional Hilbert space representation of
coherent states $|\varsigma >_{\sigma }$ to the abstract one is provided by
the following decomposition laws:

4. \textit{Vector Decomposition Law}

\begin{equation}
|\psi >=\int\limits_{%
\mathbb{C}
}d^{2}\varsigma \ |\varsigma >_{\sigma }W_{\sigma }(|\varsigma
|^{2})_{\sigma }<\varsigma |\psi >.  \label{eq35}
\end{equation}

5. \textit{Operator Decomposition Law}

\begin{equation}
\mathcal{A}=\int\limits_{%
\mathbb{C}
}d^{2}\varsigma _{1}d^{2}\varsigma _{2}\ |\varsigma _{1}>_{\sigma }W_{\sigma
}(|\varsigma _{1}|^{2})_{\sigma }<\varsigma _{1}|\mathcal{A}|\varsigma
_{2}>_{\sigma }W_{\sigma }(|\varsigma _{2}|^{2})_{\sigma }<\varsigma _{2}|.
\label{eq36}
\end{equation}

Thus, we conclude that the resolution of unity Eq.(\ref{eq8.3}) with $%
W_{\sigma }(|\varsigma |^{2})$ given by Eq.(\ref{eq8.9a}), provides an
appropriate inner product Eq.(\ref{eq30}) and allows us to introduce the
Hilbert space, Eqs.(\ref{eq33}) - (\ref{eq36}).

\section{Conclusion}

The properties of stretched coherent states were investigated. Proof that
stretched coherent states are generalized coherent states was presented. It
has been shown that stretched coherent states retain the fundamental
properties of standard coherent states and generalize the resolution of
unity or completeness condition, as well as the probability distribution
that $n$ photons are in a stretched coherent state. The stretched
displacement and stretched squeezing operators are introduced and the
multiplication law for stretched displacement operator is established.

Properties of coherent states resulting from an action of stretched
displacement and stretched squeezing operators on the vacuum state and the
Fock state are studied. Stretched squeezed stretched coherent states and
stretched squeezed stretched displaced number states were introduced and
their properties were studied.

The inner product of two quantum mechanical vectors was defined in terms of
their stretched coherent state representations, and a functional Hilbert
space was introduced.

The presented new concepts of quantum optics include two parameters, $\sigma 
$, $0<\sigma \leq 1$ and $\upsilon $, $0<\upsilon \leq 1$. In the limiting
case, when $\sigma =1$ and $\upsilon =1$, all our new results turn into the
known equations of the theory of standard coherent states.

\end{document}